\documentclass[prd,twocolumn,superscriptaddress]{revtex4}
\usepackage{adjustbox}
\usepackage{amsthm,amsmath,amssymb}
\usepackage{mathrsfs}
\usepackage{graphicx}
\usepackage{amsfonts}
\usepackage{color}
\usepackage{amscd}
\usepackage{bm}
\usepackage{cancel}
\makeatletter
\renewcommand{\maketag@@@}[1]{\hbox{\m@th\normalsize\normalfont#1}}%
\makeatother
\begin{document}
\title{Extended black hole thermodynamics from extended Iyer--Wald formalism}

\author{Yong Xiao}
\email{xiaoyong@hbu.edu.cn}
\affiliation{Key Laboratory of High-precision Computation and Application of Quantum Field Theory of Hebei Province,
College of Physical Science and Technology, Hebei University, Baoding 071002, China}
\affiliation{Hebei Research Center of the Basic Discpline for Computational Physics, Baoding, 071002, China}

\author{Yu Tian}
\email{ytian@ucas.ac.cn}
\affiliation{School of Physics, University of Chinese Academy of Sciences, Beijing 100049, China}
\affiliation{Institute of Theoretical Physics, Chinese Academy of Sciences, Beijing 100190, China}

\author{Yu-Xiao Liu}
\email{liuyx@lzu.edu.cn}
\affiliation{Key Laboratory for Quantum Theory and Applications of the Ministry of Education, Lanzhou Center for Theoretical Physics, Lanzhou University, Lanzhou, Gansu 730000, China}

\begin{abstract}
In recent years, there has been significant interest in the field of extended black hole thermodynamics, where the cosmological constant and/or other coupling parameters are treated as thermodynamic variables. Drawing inspiration from the Iyer-Wald formalism, which reveals the intrinsic and universal structure of conventional black hole thermodynamics, we illustrate that a proper extension of this formalism also unveils the underlying theoretical structure of extended black hole thermodynamics. As a remarkable consequence, for any gravitational theory described by a diffeomorphism invariant action, it is always possible to construct a consistent extended thermodynamics using this extended formalism.
\end{abstract}

\maketitle

\section{Introduction}Black hole thermodynamics plays a central role in understanding the characteristics of quantum gravity. In recent years, the so-called extended black hole thermodynamics has been proposed and substantially developed \cite{Kastor:2009wy,Dolan:2010ha,Kubiznak:2012wp}. For instance, by treating the cosmological constant $\Lambda$ as a thermodynamic variable, the first law of thermodynamics of the AdS-Schwarzschild black hole can be expressed as
\begin{align}
    \tilde{\delta} M=T \tilde{\delta} S + V \tilde{\delta} P,\label{1steq}
\end{align}
where $P\equiv -\frac{\Lambda}{8\pi}$ and $V=\frac{4\pi}{3}r_h^3$ are commonly referred to as the pressure and thermodynamic volume in the literature. However, literally interpreting $\Lambda$ as pressure could be misleading in some sense \footnote{If one were to literally interpret $\Lambda$ as pressure and compare eq.\eqref{1steq} with the standard thermodynamic law $d(\text{enthalpy})=TdS+\mathcal{V} dp$, one might suggest the black hole mass as the enthalpy rather than the energy. But black hole mass has a natural explanation as energy since it is the conserved charge associated with the Killing vector $\xi_t$.}. A better strategy is to put $\Lambda$ on equal footing with other couplings in the action. Note that we use $\tilde{\delta}$ to represent variations in the extended thermodynamics, distinguishing it from the conventional variation $\delta$ that satisfies the thermodynamic law $\delta M=T \delta S$.

At first glance, this extension may seem perplexing, as $\Lambda$ was traditionally considered a fixed parameter in the theory. However, as suggested in Refs.\cite{Cvetic:2010jb,Meessen:2022hcg}, the cosmological constant could become a dynamical variable in gauged supergravity and string theories. And a latest work \cite{Frassino:2022zaz} illustrated that the variation of the cosmological constant could be induced by tuning the brane tension in a brane-world model. Therefore, it is generally possible to circumvent the problem by taking $\Lambda$ as an external parameter controlled by a more comprehensive theory.

The extended black hole thermodynamics has offered a multitude of intriguing physical implications and applications. It serves as a fertile ground for investigating black hole phase transitions, triggering the emergence of a research direction known as black hole chemistry \cite{Kubiznak:2016qmn,Wei:2015iwa,Wei:2019uqg}. Moreover, within the framework of AdS/CFT, the holographic dual of the extended black hole thermodynamics on the CFT side has been widely explored. The AdS/CFT correspondence dictates that $c\propto \frac{l^{D-2}}{G_N}$, so the cosmological constant is not directly related to the CFT pressure but rather to the central charge; see \cite{Cong:2021fnf,Ahmed:2023snm} for details. Besides, the extended thermodynamics can also find applications in the study of holographic complexity, weak cosmic censorship conjecture, and other research fields \cite{AlBalushi:2020rqe,Gwak:2017kkt,Harlow:2022ich}.

In fact, it is not only the cosmological constant that can be considered as a thermodynamic variable; other coupling parameters in higher curvature theories of gravity, such as those in Lovelock gravity, can also be regarded as thermodynamic variables \cite{Dutta:2022wbh,Kastor:2010gq,Frassino:2014pha,Dolan:2014vba,Sinamuli:2017rhp}. Given the successes and universality of the extended thermodynamics, a natural question arises as to whether there exists a fundamental theoretical framework that underlies it. We will show that the answer is affirmative.

In this paper, we propose a robust formalism that guarantees the presence of the extended thermodynamic law for any diffeomorphism invariant theory of gravity. More importantly, it also provides a novel and systematic approach for computing the conjugate quantities associated with the couplings. In the past, the expression for the thermodynamic volume was primarily derived through the thermodynamic relation $ V^{th}\equiv (\frac{\partial M}{\partial P})_{S,J,\cdots}$ itself. In contrast, our formalism enables an independent calculation of such quantities, which has unexpectedly resolved an ambiguity between the thermodynamic volume and the geometric volume present in the literature. All the detailed calculations are provided in a supplemental Mathematica Notebook.

\vspace{0.15cm}

\section{Iyer--Wald formalism for conventional black hole thermodynamics}As is well known, for any diffeomorphism invariant gravitational theory, the conventional black hole thermodynamic law can be easily obtained by employing the Iyer--Wald formalism \cite{Wald:1993nt,Iyer:1994ys,Compere:2018aar}.

Consider the Lagrangian denoted by $\mathbf{L}=L \boldsymbol{\epsilon}$, where $\boldsymbol{\epsilon}$ is the volume form of the $D=d+1$ dimensional space-time. Its variation with respect to the dynamic fields $\phi\equiv \{g_{\mu \nu},\psi\}$ can be expressed as
\begin{align}
    \delta \mathbf{L}=\mathbf{E}^{\phi}\delta  \phi+ d \boldsymbol{\Theta}[\delta \phi], \label{gelvariform}
\end{align}
where $\mathbf{E}^{\phi}$ represents the equations of motion of the fields, and $d \boldsymbol{\Theta}[\delta \phi]$ is a total derivative term. The differential forms on space-time are written in boldface, and below we implicitly use the on-shell condition $\mathbf{E}^{\phi}=0$.

For an arbitrary, fixed vector $\xi$, the familiar Noether current $\mathbf{J}_\xi$ is defined as
\begin{align}
    \mathbf{J}_\xi \equiv \boldsymbol{\Theta}[\mathcal{L}_\xi \phi]-\xi  \cdot \mathbf{L}.
    \label{gelj}
\end{align}
Under the on-shell condition, $d  \mathbf{J_\xi}=0$, so the Noether charge $\mathbf{Q}_\xi$ can be constructed from $\mathbf{J}_\xi= d \mathbf{Q}_\xi$. Then, requiring $\xi$ to be Killing, two pivotal formulas can be proven
\begin{align}
    & d( \delta  \mathbf{Q}_\xi  -  \xi\cdot \boldsymbol{\Theta}[\delta \phi]) =0, \label{formula1}\\
     & d \mathbf{Q}_\xi = -\xi  \cdot \mathbf{L}.
     \label{formula2}
\end{align}
The proof of eq.(\ref{formula1}) has been left to the Appendix; eq.(\ref{formula2}) is a direct consequence of the Killing property $\mathcal{L}_\xi \phi=0$.

For a stationary black hole with a
bifurcate Killing horizon, we integrate the two formulas over a hypersurface $\mathcal{V}_r$ extending from the bifurcation surface denoted by $S_h$ to another codimension-2 surface $S_r$. Owing to Gauss's theorem, eqs.\eqref{formula1} and \eqref{formula2} become
\begin{align}
 \int_{S_r} &\left(  \delta \mathbf{Q}_{\xi_H} -  \xi_H \cdot \boldsymbol{\Theta}[\delta \phi] \right) - \int_{S_h} \delta \mathbf{Q}_{\xi_H}    =0, \label{intfor1} \\
  & \int_{S_r} \mathbf{Q}_{\xi_H}-\int_{S_h} \mathbf{Q}_{\xi_H} = -\int_{\mathcal{V}_r}\xi_H  \cdot \mathbf{L}, \label{intfor2}
\end{align}
where we have applied the horizon Killing vector $\xi_H$ that vanishes on $S_h$. Due to the unique characteristics of the horizon Killing vector and the bifurcation surface, one can show that $ \int_{S_h} \delta \mathbf{Q}_{\xi_H}=T\delta S$ and similarly $\int_{S_h} \mathbf{Q}_{\xi_H} = TS$ for gravity sector, which actually is the origin of the definition of Wald entropy \cite{Wald:1993nt,Iyer:1994ys}. Thus it is essential to use $\xi_H$ in order to establish a connection with the thermodynamic properties of a black hole. Thereupon, the first law of black hole thermodynamics follows from eq.\eqref{intfor1}, and the Smarr relation follows from eq.\eqref{intfor2}.

As an example, for a Kerr black hole in $D=4$ dimensions, the horizon Killing vector is given by $\xi_H=\frac{\partial}{\partial t}+\Omega_H \frac{\partial}{\partial \phi}$. In eq.\eqref{intfor1}, the term $\int_{S_r} \left(  \delta \mathbf{Q}_{\xi_H} -  \xi_H \cdot \boldsymbol{\Theta}[\delta \phi] \right) $ reduces to $\delta M-\Omega_H \delta J$, leading to the first law $\delta M-\Omega_H \delta J-T\delta S=0$. Meanwhile, eq.\eqref{intfor2} becomes $(\frac{M}{2}-\Omega_H J)-TS=0$, by noticing $\int_{\mathcal{V}_r}\xi  \cdot \mathbf{L}$ vanishes on-shell for the Kerr solution. This yields the Smarr relation $M=2TS+2\Omega_H J$.

We emphasize that eqs.\eqref{intfor1} and \eqref{intfor2} are valid for any surface $S_r$ of radius $r$ that encompasses the black hole, which is not necessarily the spatial infinity. In realistic calculations, all the terms dependent on $r$ will cancel out in the final result, as a distinctive feature of the covariant formalism. Thus we don't have to worry about the potential divergent behaviors caused by $r\rightarrow \infty$, particularly in the context of AdS black holes.

\vspace{0.15cm}

\section{An extension of the Iyer--Wald formalism}Let's consider a general diffeomorphism invariant gravitational theory with the Lagrangian
\begin{align}
L= \frac{1}{16\pi} (R-2\Lambda ) + \sum_m \alpha_m F_m [g_{\mu\nu},\nabla_\rho, R_{\mu \nu \rho \sigma}],\label{gelactionla}
\end{align}
where $F_m$ represents the higher curvature term. In the spirit of effective field theory, it is permissible to add all possible diffeomorphism invariant curvature terms to the Einstein-Hilbert action \cite{Cardoso:2018ptl,Dutta:2022wbh}. Without confusion, we also use the subscript $m$ to count the number of the curvatures and half of the number of $\nabla_\rho$'s in $F_m$, so that the coupling parameter $\alpha_m$ has the dimension $[L]^{2(m-1)}$. We neglect the matter field sectors, thus $\phi=g_{\mu \nu}$ is the only dynamical field.

We introduce the operator $\tilde{\delta}$ which allows for variations in the cosmological constant $\Lambda$ and other couplings $\alpha_m$, in contrast to the conventional variation $\delta$. Technically, we define $\tilde\delta g_{\mu \nu}=\delta g_{\mu \nu}+\frac{\partial g_{\mu \nu}}{\partial \Lambda}\tilde\delta \Lambda + \sum_m \frac{\partial g_{\mu \nu}}{\partial \alpha_m}\tilde\delta \alpha_m$, which naturally arises in analyzing the complete variation $\tilde \delta \mathbf{L}(g_{\mu \nu}, \Lambda,\alpha_m)$. First, $\tilde \delta \mathbf{L}$ includes a term $\frac{\partial \mathbf{L}}{\partial g_{\mu\nu}}\delta g_{\mu\nu} $ induced by varying $g_{\mu \nu}$ while fixing $\Lambda$ and $\alpha_m$'s. Second, it includes the terms $\frac{\partial \mathbf{L}}{\partial g_{\mu\nu}} \frac{\partial g_{\mu \nu}}{\partial \Lambda} \tilde\delta \Lambda +\frac{\partial \mathbf{L}}{\partial \Lambda} \tilde\delta \Lambda $ induced by varying $\Lambda$, either from the explicit dependence of $\mathbf{L}$ on $\Lambda$, or from implicit dependence of $\mathbf{L}$ on $\Lambda$ through $g_{\mu\nu}$. And $\tilde \delta \mathbf{L}$ also includes similar terms caused by varying $\tilde \delta \alpha_m$. Adding them together, one easily identifies the presence of $\frac{\partial \mathbf{L}}{\partial g_{\mu\nu}} \tilde \delta g_{\mu \nu}$. Accordingly, the complete variation is
\begin{align}
\begin{split}
     \tilde{\delta} \mathbf{L}=& \frac{\partial \mathbf{L}}{\partial g_{\mu\nu}}\tilde{\delta} g_{\mu\nu}+ \frac{\partial \mathbf{L}}{\partial \Lambda}\tilde\delta \Lambda + \sum_m \frac{\partial \mathbf{L}}{\partial \alpha_m}\tilde\delta \alpha_m\\
     =&\mathbf{E}^{\phi} \tilde{\delta}  \phi+  d \boldsymbol{\Theta}[\tilde{\delta} \phi]-\frac{\tilde{\delta} \Lambda}{8\pi}\boldsymbol{\epsilon}  + \sum_m F_m  \boldsymbol{\epsilon} \, \tilde{\delta} \alpha_m.
\end{split}
\end{align}
Utilizing this expression of $\tilde\delta \mathbf{L}$, we reexamine the derivation of eq.\eqref{formula1} as given in the Appendix. It leads to
\begin{align}
d(\tilde{\delta} \mathbf{Q}_\xi -  \xi\cdot \boldsymbol{\Theta}[\tilde{\delta} \phi]) =\left( \frac{\tilde{\delta} \Lambda}{8\pi}  - \sum_m F_m \tilde{\delta} \alpha_m  \right) \xi\cdot\boldsymbol{\epsilon}  .
\end{align}
By applying the horizon Killing vector $\xi_H$ and integrating it over the hypersurface $\mathcal{V}_r$, we obtain
\begin{align}
\begin{split}
   & \int_{S_r}\left(\tilde{\delta} \mathbf{Q}_{\xi_H} -  {\xi_H}\cdot \boldsymbol{\Theta}[\tilde{\delta} \phi] \right) - \int_{S_h} \tilde{\delta} \mathbf{Q}_{\xi_H}  \\ = & \frac{\tilde{\delta} \Lambda}{8\pi}\, \int_{\mathcal{V}_r}{\xi_H}\cdot\boldsymbol{\epsilon} - \sum_m\tilde{\delta} \alpha_m  \int_{\mathcal{V}_r} F_m \, {\xi_H}\cdot\boldsymbol{\epsilon}.
   \end{split}
\label{myformula1}
\end{align}
This formula yields the extended first law of black hole thermodynamics, similar to how eq.\eqref{intfor1} produces the conventional first law. As mentioned, in realistic calculations all the terms dependent on $r$ (including potentially divergent terms) precisely cancel out, producing the desired result.

Since we are mainly concerned about the AdS black holes, we also introduce an alternative method to regularize the divergences, which involves subtracting the contribution of the pure AdS as the regulator. For a pure AdS background, eq.\eqref{myformula1} still holds by omitting the horizon term. By subtracting it from eq.\eqref{myformula1}, the divergent terms can be eliminated. Then taking the limit $r\rightarrow \infty$ removes the other $r$-dependent terms. This procedure defines the regularized integral $\int^{(reg.)}
\equiv \int^{(BH)}-\int^{(AdS)}$ and leads to the regularized version of eqs.\eqref{myformula1} as
\begin{align}
\begin{adjustbox}{width=\linewidth-2.8em}
$ \displaystyle
\int_{S_\infty}^{(reg.)}\!\left(\tilde{\delta} \mathbf{Q}_{\xi_H} \! -\! \xi_H\! \cdot\! \boldsymbol{\Theta}[\tilde{\delta} \phi] \right)\! -\! T\tilde\delta S \!=\! V \tilde{\delta}P\!+\! \sum\limits_m V_m \tilde{\delta} \alpha_m,
$
\end{adjustbox}
\label{myformulareg}
\end{align}
where $P\equiv -\frac{\Lambda}{8\pi}$, the horizon term $\int_{S_h}\tilde{\delta}\mathbf{Q}_{\xi_H}$ has been formally identified as $T \tilde{\delta} S $, and the geometric volume $V$ and its generalizations $V_m$ are defined respectively as
\begin{align}
   V\equiv & -\int_\mathcal{V_\infty}^{(reg.)} \xi_H \cdot \boldsymbol{\epsilon}, \label{vformula} \\
 V_m\equiv & -\int_\mathcal{V_\infty}^{(reg.)} F_m \, \xi_H \cdot \boldsymbol{\epsilon}.
\end{align}
Amazingly, eq.\eqref{myformulareg} has almost been the form of the extended first law. We solely need to evaluate the first term of eq.\eqref{myformulareg}. Notice that the background subtraction can also be applied in the analysis of conventional thermodynamics. Anyway, the conventional case can be regarded as a special case of the extended formalism with $\tilde\delta P=0$ and $\tilde\delta \alpha_m=0$.

From our derivation, it is clear that the term $V\tilde \delta P$ in eq.\eqref{myformulareg} comes from the explicit dependence of the Lagrangian on the cosmological constant $\Lambda$. Meanwhile, during the evaluation of $\int_{S_\infty}^{(reg.)}\left(\tilde{\delta} \mathbf{Q}_{\xi_H} -  \xi_H\cdot \boldsymbol{\Theta}[\tilde{\delta} \phi] \right)$, additional terms proportional to $\tilde\delta P$, denoted by $\Delta V \, \tilde\delta P$, may also arise due to the dependence of the dynamical fields $\phi$ on $\Lambda$. Combining them together, we will get the thermodynamic volume $V^{th}=V+\Delta V$, which is consistent with the long-lasting experience that the thermodynamic volume $V^{th}$ is not necessarily equal to the geometric volume $V$. From this perspective, $V\tilde \delta P$ is just a normal term, while the additional terms are the truly interesting ones, as they explain the distinction between $V^{th}$ and $V$. In a similar manner, the conjugate quantities $V^{th}_m=V_m+\Delta V_m $ of $\alpha_m$'s can also be deduced.

As before, the Smarr relation can be derived from eq.\eqref{intfor2}. This formula remains unchanged in the discussion of the extended black hole thermodynamics, since it is evaluated on a given metric and has nothing to do with the variations. However, a scaling argument combined with the extended first law is sufficient to deduce the Smarr relation. Therefore, we can make use of eq.\eqref{intfor2} for cross-checking.
\vspace{0.15cm}

\section{AdS-Schwarzschild black hole}Now we start to study some concrete examples, which are limited to 4 dimensions for simplicity. In Einstein gravity with a negative cosmological constant $\Lambda$, the Lagrangian is $L=\frac{1}{16\pi}(R-2\Lambda)$. From the variation $\delta ( \sqrt{-g} L )= \sqrt{-g} E^{\mu \nu}\delta g_{\mu\nu} + \sqrt{-g} \nabla_\mu \Theta^\mu$, one can read off
\begin{align}
\Theta^\mu [\delta g_{\mu \nu}]\equiv \frac{1}{16 \pi} (g^{\mu \alpha} \nabla^\nu \delta g_{\alpha \nu} - g^{ \alpha \beta} \nabla^{\mu} \delta g_{ \alpha \beta} ), \label{einsteintheta}
\end{align}
and construct the Noether charge as
\begin{align}
    Q_\xi^{\mu\nu} \equiv -\frac{1}{16\pi}  \left (  \nabla^\mu \xi^\nu -\nabla^{\nu}\xi^\mu  \right ). \label{einq}
\end{align}
Their differential forms are given by $\boldsymbol{\Theta}\equiv\frac{1}{3!}\Theta^{\mu}\epsilon_{\mu \nu \alpha \beta} dx^\nu \wedge dx^\alpha \wedge dx^\beta$, and $\mathbf{Q}_\xi=\frac{1}{2! 2!}Q_\xi^{\mu \nu}\epsilon_{\mu \nu \alpha \beta} dx^\alpha \wedge dx^\beta$.

For the AdS-Schwarzschild balck hole, the horizon Killing vector is $\xi_H=\frac{\partial}{\partial t}$ and the horizon radius is denoted by $r_h$. Substituting eqs.\eqref{einsteintheta} and \eqref{einq} into the unregularized formula \eqref{myformula1}, we get
\begin{align}
    \tilde{\delta} M+\cancel{\frac{r^3}{6}\tilde{\delta} \Lambda} -T\tilde{\delta} S=\cancel{\frac{4 \pi r^3}{3} \frac{\tilde{\delta} \Lambda }{8\pi} }-  \frac{4 \pi {r_h}^3}{3} \frac{\tilde{\delta} \Lambda }{8\pi}.
\end{align}
Alternatively, the regularized formula \eqref{myformulareg} also yields
\begin{align}
    \tilde\delta M=T\tilde\delta S +V \tilde\delta P,
    \label{ads_bh1law}
\end{align}
with $ V=\frac{4 \pi}{3} r_h^3$. Both approaches produce the same final result, but the latter approach is more straightforward. So we will mainly employ the regularized formulas. Note that there are no additional terms proportional to $\tilde\delta P$ in this case, thus the geometrical volume $V$ and the thermodynamic volume $V^{th}$ coincide with each other.

The extended first law, together with a scaling argument, can lead to the Smarr relation. Here the scaling behavior is given by $\eta M(S,P)=M (\eta^2 S,\eta^{-2}P)$. Taking the derivative with respect to $\eta$, one gets $ M=2\frac{\partial{M}}{\partial{S}}S-2\frac{\partial{M}}{\partial{P}}P$. By virtue of eq.\eqref{ads_bh1law}, it becomes
\begin{align}
    M=2TS-2PV.
\end{align}
Note this Smarr relation can also be derived from eq.\eqref{intfor2}.

\vspace{0.15cm}

\section{AdS-Kerr black hole}The AdS-Kerr black hole is a nontrivial example that deserves detailed inspection. The metric is listed in the Appendix, parameterized by the mass parameter $m$, rotational parameter $a$ and AdS radius $l$. In fact, there were some controversies even around the conventional first law of thermodynamics for the AdS-Kerr case; see \cite{Gibbons:2004ai,Caldarelli:1999xj} for details. Nevertheless, both the conventional first law and its extension can be naturally derived in the present formalism.

Here, the horizon Killing vector is $\xi_H=\frac{\partial}{\partial t}+\Omega_H \frac{\partial}{\partial \phi} $. The AdS-Kerr spacetime has a nonvanishing angular velocity at infinity, that is, $\Omega_\infty=-\frac{a}{l^2}$. Thus, we can make the decomposition $\xi_H=\xi_t+\Omega\, \xi_\phi$, where $\xi_t \equiv \frac{\partial}{\partial t}+\Omega_{\infty} \frac{\partial}{\partial \phi}$,  $\xi_\phi\equiv \frac{\partial}{\partial \phi}$ and  $\Omega\equiv \Omega_H-\Omega_\infty$. By the linearity property of the formalism, the two parts of the horizon Killing vector can be analyzed separately. First, we have $\int^{(reg.)}_{S_\infty} \left(  \delta \mathbf{Q}_{\xi_t} -  \xi_t \cdot \boldsymbol{\Theta}[\delta \phi] \right)= \delta M$, where $M= \frac{m}{(1-\frac{a^2}{l^2})^2}$. Second, by noticing $\xi_\phi$ is tangential to the integrating surface, there is $ \int^{(reg.)}_{S_\infty}  \left(  \delta \mathbf{Q}_{\xi_\phi} -  \xi_\phi \cdot \boldsymbol{\Theta}[\delta \phi] \right)=\delta\,\int^{(reg.)}_{S_\infty}
 \mathbf{Q}_{\xi_\phi}= -\delta J$,  where $J=\frac{m a}{(1-\frac{a^2}{l^2})^2}$. Therefore we get the conventional first law of thermodynamics
\begin{align}
    \delta M-\Omega\, \delta J- T\,\delta S=0.
\end{align}
Notably, the angular velocity that appears in the black hole thermodynamics is $\Omega$ rather than $\Omega_H$.

Next we derive the extended black hole thermodynamics from the formula \eqref{myformulareg}. When evaluating  $\int^{(reg.)}_{S_\infty}  \left(  \tilde\delta \mathbf{Q}_{\xi_\phi} -  \xi_\phi \cdot \boldsymbol{\Theta}[\tilde\delta \phi] \right)$,  because $\xi_\phi$ is tangential to the integrating surface, its second term vanishes. This leads to a total variation $ \tilde\delta \int^{(reg.)}_{S_\infty}
 \mathbf{Q}_{\xi_\phi}$, which is identified as $ -\tilde\delta J$ by definition. Then we have to explicitly calculate  $ \int^{(reg.)}_{S_\infty}  \left(  \tilde\delta \mathbf{Q}_{\xi_t} -  \xi_t \cdot \boldsymbol{\Theta}[\tilde\delta \phi] \right)$ associated with the Killing vector $\xi_t$, which gives
\begin{align}
    \frac{l^4}{\left(l^2-a^2\right)^2} \tilde\delta m+\frac{4 a l^4 m}{\left(l^2-a^2\right)^3} \tilde\delta a-\frac{a^2 l m \left(a^2+3 l^2\right)}{\left(l^2-a^2\right)^3} \tilde\delta l.\label{noninte}
\end{align}
Unlike the conventional case analyzed above, now the expression is nonintegrable. By the knowledge of calculus, if we extract a total variation $\tilde\delta M$ from the nonintegrable expression, there \emph{inevitably} remains an additional term. Indeed, eq.\eqref{noninte} turns out to be $\tilde{\delta} M-\frac{4\pi}{3}Ma^2 \tilde{\delta} P$,
where $ P\equiv -\frac{\Lambda}{8\pi} = \frac{3}{8\pi} \frac{1}{l^2}$. Accordingly, we find the extended first law of thermodynamics
\begin{align}
    \tilde{\delta} M=T\,\tilde{\delta} S + \Omega \,\tilde{\delta} J+ \left(V + \frac{4\pi}{3}Ma^2 \right) \,\tilde{\delta} P. \label{kerrext}
\end{align}
Thus, in the AdS-Kerr case, the thermodynamic volume $V^{th}= V + \frac{4\pi}{3}Ma^2  $ isn't equal to the geometric volume $V$. This result \eqref{kerrext} coincides with that given in \cite{Cvetic:2010jb}, which was determined from thermodynamic relations. In stark contrast, in our formalism, the thermodynamic volume $V^{th}$ and the first law are deduced even \emph{without} resorting to the specific expressions of $T$, $S$, $\Omega$, $J$ and $V$.

Taking the derivative of the scaling behavior $\eta M(S,J,P)=M (\eta^2 S,\eta^2 J,\eta^{-2}P)$ with respect to $\eta$, one gets $M=2\frac{\partial M}{\partial S} S+2\frac{\partial M}{\partial J} J-2\frac{\partial M}{\partial P}P$. Using the extended first law \eqref{kerrext}, one can read off the Smarr relation
\begin{align}
    M=2TS+2\Omega J-2 \left(V + \frac{4\pi}{3}Ma^2 \right)P. \label{kerrsmarr}
\end{align}
Now let's rederive this Smarr relation from eq.\eqref{intfor2} for cross-checking. Applying $\xi_H=\xi_t+\Omega \,\xi_\phi$, the regularized version of eq.\eqref{intfor2} can be put into the form
\begin{align}
\begin{adjustbox}{width=\linewidth-2.8em}
$ \displaystyle
 2 \! \int_{S_{\infty}}^{(reg.)} \mathbf{Q}_{\xi_t}\!+\! 2 \Omega \! \int_{S_{\infty}}^{(reg.)} \mathbf{Q}_{\xi_\phi} \!-\!2 \int_{S_h} \mathbf{Q}_{\xi_H} \!=\! - \! 2 \! \int_{\mathcal{V}_{\infty}}^{(reg.)}\xi_H \! \cdot \!\mathbf{L}.
$
\end{adjustbox}\label{smarrreg}
\end{align}
 We have multiplied a factor $2$, because now the first term is just the standard Komar mass formula, which is regularized for the the case of AdS black hole. By explicit calculation, we find
 \begin{align}
    M_K\equiv 2 \int_{S_{\infty}}^{(reg.)} \mathbf{Q}_{\xi_t}=M+\frac{8\pi}{3}Ma^2 P,\label{kerrmkm}
\end{align}
and $ -2 \int_{\mathcal{V}_{\infty}}^{(reg.)}\xi_H  \cdot \mathbf{L}=-2PV $. Substituting them into eq.\eqref{smarrreg}, it reduces to $(M+\frac{8\pi}{3}Ma^2 P)  -2 \Omega J -2 TS=-2PV$, which surely recovers the Smarr relation \eqref{kerrsmarr}.

Now we can identify the presence of additional terms $\Delta V$ in the first law \eqref{kerrext} and $\Delta M$ in the Smarr relation \eqref{kerrsmarr}. Specifically, $\Delta M\equiv M-M_K$ is defined as the difference between the canonical energy and the Komar energy, while $\Delta V\equiv V^{th}-V$ is defined as the difference between the thermal volume and the geometric volume. They play an essential role in the extended thermodynamics, i.e., unless $\Delta M$ and $\Delta V$ have been appropriately taken into account, a well-defined Smarr relation and extended first law cannot be achieved.

We notice that there have been some efforts to address the extended first law from some general formalisms \cite{Kubiznak:2016qmn,Jacobson:2018ahi, Bhattacharya:2017hfj,Wu:2016auq}. But, as far as we know, there was no similar analysis for the AdS-Kerr case that accurately reproduces eq.\eqref{kerrext}. As we stressed earlier, the additional terms are truly nontrivial and interesting. Therefore, to distinguish our approach from others, below we make further discussions about how these terms could emerge within our formalism.

Firstly, below eq.\eqref{noninte}, we have exemplified how the additional term $\Delta V$ arises from a re-examination about the integrability condition. Acting $\tilde{\delta}$ on a physical quantity such as $M$ requires an extra derivative term $\frac{\partial M}{\partial \Lambda}\tilde\delta \Lambda$, compared to $\delta M$. One must take this into account in constructing the total variation $\tilde\delta M$. Even if the conventional expression $\int^{(reg.)}_{S_\infty} \left(  \delta \mathbf{Q}_{\xi_t} -  \xi_t\cdot \boldsymbol{\Theta}[\delta \phi] \right) $ can be identified as $\delta M $ under integrability condition, the extended expression $ \int^{(reg.)}_{S_\infty} \left( \tilde{\delta} \mathbf{Q}_{\xi_t} -  \xi_t\cdot \boldsymbol{\Theta}[\tilde{\delta} \phi] \right) $ may become nonintegrable, thus it can not be naively identified as $\tilde{\delta} M$. This point is very crucial, yet it can be easily ignored.

Secondly, the additional terms $\Delta V$ and $\Delta M$ reflect the nontrivial asymptotic behaviors of the AdS-Kerr case. These additional terms are actually the terms that survive in the limit $r\rightarrow \infty$, when evaluating the corresponding formulas such as eqs.\eqref{myformulareg} and \eqref{smarrreg}, in addition to the regular terms. Moreover, our analysis reveals a close correlation between $\Delta M $ and $\Delta V$, given by $\Delta M= -2 P \, \Delta V$. Soon we will see that it can be viewed as a special case of eq.\eqref{triangleidentity} with $n=0$. Interestingly, although the situation $\Delta M \neq 0$ may appear problematic for someone attempting to calculate the canonical energy $M$ from the Komar mass formula, the difference is essential in the context of the extended black hole thermodynamics.

\vspace{0.15cm}

\section{Asymptotically-AdS black hole in higher curvature gravity}
The extended formalism is equally applicable to higher curvature theories of gravity, where the coupling parameters $\alpha_m$ are treated as thermodynamic variables. We introduce a new subscript $n=\{0,m\}$ and denote $P=-\frac{\Lambda}{8\pi}=\alpha_0$, $F_0=1$ as well as $V_0=V$, so that our results could be put into a unified form. We make general analysis, and then study a concrete example.

The variation of the Lagrangian \eqref{gelactionla} leads to modifications to the Einstein field equation, which could be rather complicated. Whatever, for any higher curvature terms included in the action, it is standard to take the couplings $\alpha_m$ as small quantities, and solve for black hole solutions perturbatively and iteratively \cite{Cardoso:2018ptl}.

For simplicity, we restrict our analysis to the static and spherically symmetric black hole solutions, where the horizon Killing vector is simply $\xi_t=\frac{\partial}{\partial t}$. The expressions of $\Theta^\mu$ and $Q^{\mu\nu}$ for the higher curvature theory can be obtained using standard techniques \cite{Iyer:1994ys}. Subsequently, we evaluate the integral $ \int^{(reg.)}_{S_\infty} \left( \tilde{\delta} \mathbf{Q}_{\xi_t} -  \xi_t\cdot \boldsymbol{\Theta}[\tilde{\delta} \phi] \right) $ in eq.\eqref{myformulareg}. In general, we expect it reduces to
the form $ \tilde{\delta} M- \sum_n\Delta V_n\, \tilde{\delta} \alpha_n$, where $\Delta V_n\, \tilde{\delta} \alpha_n$ represents the additional term proportional to $\tilde{\delta} \alpha_n$. Accordingly,  eq.\eqref{myformulareg} simplifies to a general form of the extended first law
\begin{align}
    \tilde{\delta}M=T\,\tilde{\delta} S+\sum_n (V_n+\Delta V_n)\,\tilde{\delta}\alpha_n.
\end{align}

From the extended first law and the scaling behavior $\eta M(S,\cdots,\alpha_n)=M (\eta^{2} S,\cdots,\eta^{2(n-1)}a_n)$, we deduce the Smarr relation
\begin{align}
    M=2TS + 2 \sum_n (n-1)  \left(V_n+\Delta V_n\right)\alpha_n. \label{smarrhigher}
\end{align}
Let's rederive the Smarr relation from eq.\eqref{kerrsmarr} with $\Omega=0$. It reduces to $M_K-2 T S=-2\int^{(reg.)}_{\mathcal{V}_\infty}\xi_t\cdot \boldsymbol{L}$. Because $M_K$ is not necessarily equal to the canonical energy $M$, we represent it as $M_K=M-\Delta M$. Next, we handle with $-2\int^{(reg.)}_{\mathcal{V}_\infty}\xi_t\cdot \boldsymbol{L}=-2\int^{(reg.)}_{\mathcal{V}_\infty}d^3 x\sqrt{-g}\left( \frac{R}{16\pi}+\sum_n  \alpha_n F_n \right)$, which involves the sum of the on-shell integrals of the Einstein--Hilbert term and the higher curvature terms. Interestingly, there exists a formula that establishes a relationship among such integrals \cite{Xiao:2022auy}. For an asymptotically-AdS black hole, the formula can be generalized as
\begin{align}
    \int \frac{R}{16\pi}= \sum_n  (n-2) \int \alpha_n F_n,\label{rfnrelation}
\end{align}
where $\int$ is an abbreviation of $\int^{(reg.)}_{\mathcal{V}_\infty}d^3 x \sqrt{-g}$ \footnote{Actually, the formula can be recast into an elegant form $\sum_i (i-2)\int \alpha_i F_i=0$ (or $\sum_i \frac{2i-D}{D-2}\int \alpha_i F_i=0$ in $D$ dimensions), when the Lagrangian is written as $L=\sum_i \alpha_i F_i$ with $\alpha_1 F_1$ representing $\frac{R}{16\pi}$. Thus the terms $\int \alpha_i F_i$ are not independent with each other. This explains why we do not regard $\alpha_1$ as a thermodynamic variable, since it does not present in the Smarr relation after being expressed by other $\alpha_n$'s.}. Thus we have $ - 2 \int_\mathcal{V_\infty}^{(reg.)} \xi_t \cdot \mathbf{L}=-2\sum_n (n-1)\int \alpha_n F_n  =2\sum_n (n-1)\alpha_n V_n$. This leads to
\begin{align}
    M=2TS+\Delta M+2\sum_{n} (n-1)\, \alpha_n V_n.
\end{align}
Comparing with eq.\eqref{smarrhigher}, we find there must be
\begin{align}
    \Delta M=2\sum_n (n-1)\,\alpha_n \,\Delta V_n.\label{triangleidentity}
\end{align}
Once again, there exists a close correlation between the additional terms $\Delta V_n$ and $\Delta M$.

As an example, consider a model described by the Lagrangian
\begin{align}
\begin{adjustbox}{width=\linewidth-2.8em}
$ \displaystyle
L=\frac{1}{16\pi}(R-2\Lambda)+  \alpha_2 R_{\mu\nu}R^{\mu \nu}+  \alpha_3 R_{\mu\nu}^{ \ \ \rho \sigma}R_{\rho \sigma}^{\ \  \alpha \beta}R_{\alpha \beta}^{\ \  \mu \nu}.
$
\end{adjustbox}
\label{lagrangian23}
\end{align}
We have solved the black hole solution perturbatively around the AdS-Schwarzschild metric, up to $2$nd order in the couplings, as given in the Appendix. Evaluating eq.\eqref{myformulareg}, we get the extended first law in the form $\tilde{\delta} M=T\tilde{\delta} S+V \tilde{\delta} P\!+\!   ( V_2  \!+\! \Delta V_2 )  \tilde{\delta}\alpha_2 \!+\!  ( V_3   \!+\! \Delta V_3 )  \tilde{\delta}\alpha_3$,
where
\begin{align}
  &  \Delta V_2= \frac{32 }{3}  \pi  M \Lambda  + \alpha_3 \frac{8192 }{27} \pi^2   M\Lambda^3,\\
    & \Delta V_3=  \frac{32}{3} \pi   M  \Lambda^2 + \alpha_3  \frac{10240}{27} \pi^2  M \Lambda^4.
\end{align}
On the other hand, we obtain $\Delta M\equiv M-M_K =
\frac{64}{3} \pi  M \Lambda   (\alpha_2+2 \alpha_3 \Lambda )+\frac{8192}{27} \pi ^2 \Lambda ^3 M (2 \alpha_2 \alpha_3+5 \alpha_3^2 \Lambda )$. One can easily verify the equality $\Delta M= 2\alpha_2 \Delta V_2 +  4 \alpha_3 \Delta V_3 $, as alluded to in eq.\eqref{triangleidentity}.

Notice that, when the equation of motion has been modified by the higher curvature terms, the pure AdS solution may possess an effective cosmological constant $\Lambda_{e}$ that deviates from $\Lambda$. In view of some motivation, one may prefer to utilize $\Lambda_{e}$ instead of $\Lambda$ as the thermodynamic variable \cite{Dutta:2022wbh}. It is straightforward to change the variables by substituting the relation $\Lambda=\Lambda(\Lambda_e,\alpha_2,\alpha_3)$ into the extended first law at hand.

\vspace{0.15cm}

\section{Concluding remarks}In this paper, we have fulfilled an elegant derivation of extended thermodynamics from the extended Iyer-Wald formalism, thereby establishing a robust foundation for the extended thermodynamics.

In the field of extended black hole thermodynamics,
an interesting observation is that the thermodynamic volume $V^{th}$ may not necessarily be equal to the geometric volume $V$. This observation is naturally explained within our formalism. We have shown that $V\,\tilde\delta P$ comes from the explicit dependence of the Lagrangian on $\Lambda$, while the evaluation of $ \int^{(reg.)}_{S_\infty} \left( \tilde{\delta} \mathbf{Q}_{\xi_t} -  \xi_t\cdot \boldsymbol{\Theta}[\tilde{\delta} \phi] \right) $ yields additional terms $\Delta V \, \tilde\delta P$ by a careful re-examination of the integrability condition. This contributes a novel way to determine $\Delta V$ and $V^{th}$, in contrast to those approaches relying on manipulating the thermodynamic relations. The argument applies equally well to the conjugate quantities $V_{m}^{th}=V_m+\Delta V_m$ for other couplings $\alpha_m$ in the theory.

In addition, we uncover a connection between the additional terms $\Delta V_n$ and $\Delta M$, as indicated in eq.\eqref{triangleidentity}. As explained, these terms capture the nontrivial asymptotic behaviours of the bulk theory. Through the dual thermodynamics, these terms and the relation among them may also be important on the CFT side, which is worthy of deep study in the future.

We have only analyzed a limited number of examples in this paper. However, the formalism is flexible and can be readily applied to more complex scenarios. This includes gravitational theories coupled with diverse matter fields, as well as gravitational theories in higher dimensions, such as Lovelock gravity \cite{Kastor:2010gq,Frassino:2014pha,Dolan:2014vba,Sinamuli:2017rhp}. In the Appendix, we present an analysis of the extended thermodynamics for Gauss-Bonnet gravity as a specific case of Lovelock gravity in $D=5$ dimensions \cite{Cai:2001dz}. It is valuable to verify our extended formalism on the exact black hole solutions in such theories, which means the formalism really works at any order of the couplings.

\begin{acknowledgments}
We would like to thank Hongbao Zhang for helpful discussions. This work was supported in part by Hebei Natural Science Foundation (NSF) with Grant No. A2021201022, and by NSF of China with Grant Nos. 11975235, 12035016, 11875151 and 12247101.
\end{acknowledgments}

\appendix
\section*{Appendix A}
In this Appendix, we provide a concise derivation of eqs.(4) and (5). The derivation is as follows
\begin{align}
\begin{split}
&   \delta \mathbf{J}_\xi -  d (\xi\cdot \boldsymbol{\Theta}[\delta \phi])\\
=&  \delta \boldsymbol{\Theta}[\mathcal{L}_{\xi} \phi]    - \xi\cdot \delta \mathbf{L}   -  d (\xi\cdot \boldsymbol{\Theta}[\delta \phi ]   )\\
=  &   \delta \boldsymbol{\Theta}[\mathcal{L}_{\xi} \phi]    - \xi\cdot d \boldsymbol{\Theta}[ \delta \phi]    -  d (\xi\cdot \boldsymbol{\Theta}[\delta \phi])\\
    =  & \delta \boldsymbol{\Theta}[\mathcal{L}_{\xi} \phi] -\mathcal{L}_{\xi} \boldsymbol{\Theta}[\delta \phi] \\  \equiv  &\boldsymbol{\omega} [\delta \phi, \mathcal{L}_{\xi} \phi].
\end{split} \label{wandj}\tag{A.1}
\end{align}
In the first step, we use the definition $\mathbf{J}_\xi=  \boldsymbol{\Theta}[\mathcal{L}_{\xi} \phi]    - \xi\cdot  \mathbf{L}  $ and that $\xi$ is fixed on the space-time. In the second step, we use $\delta \mathbf{L}= \mathbf{E}^{\phi}\delta  \phi+d\boldsymbol{\Theta}[\delta \phi]$. In the third step, the Cartan's magic formula $\mathcal{L}_\xi \boldsymbol{\Theta}=\xi\cdot d \boldsymbol{\Theta}+d (\xi\cdot \boldsymbol{\Theta})$ is used. In the covariant phase space method, $\boldsymbol{\omega} [\delta_1 \phi, \delta_2 \phi]$ is called the symplectic current and is linear and anti-symmetric in its arguments. For the Killing vector $\xi$ with the property $\mathcal{L}_{\xi} \phi=0$, $\boldsymbol{\omega} [\delta \phi, \mathcal{L}_{\xi} \phi]$ vanishes. Therefore, utilizing $\delta \mathbf{J}_\xi=\delta d\mathbf{Q}_\xi=d \delta \mathbf{Q}_\xi$, we arrive at eq.(4).

In the above derivation, we implicitly employed the on-shell condition $ \mathbf{E}^{\phi}=0$ and, even more, $\delta \mathbf{E}^\phi=0$. The latter condition implies that the variation $\delta$ in eq.(4) is not arbitrary, but rather connects two neighboring solutions of the theory. If extending the variation from $\delta$ to $\tilde\delta$, we still have $\tilde\delta \mathbf{E}^\phi=0$, because the Lagrangian, the equation of motion and its solution are all varied simultaneously while adjusting $\Lambda$.

\section*{Appendix B}
In this appendix, we provide a list of the black hole metrics utilized in the main text of this paper.

The metric of the AdS-Schwarzschild black hole reads
\begin{align}
    d s^2=\!-\!\left(1\!-\!\frac{2M}{r}\!-\!\frac{ \Lambda\, r^2}{3}\right) dt^2\!+\!\frac{1}{1\!-\!\frac{2M}{r}\!-\!\frac{ \Lambda\, r^2}{3}} dr^2\!+\!r^2 d \Omega_2.
\tag{B.1}
\end{align}

The metric of the AdS-Kerr black hole in Boyer-Lindquist coordinates is given by
\begin{align}
\begin{split}
  ds^2 =& -\frac{\Delta_r}{\rho^2}\left(dt - \frac{a\sin^2\theta}{\Xi}\,d\phi\right)^2 + \frac{\rho^2}{\Delta_r}\,dr^2 + \frac{\rho^2}{\Delta_\theta}\,d\theta^2\\ & + \frac{\Delta_\theta\sin^2\theta}{\rho^2}\left(a\,dt - \frac{r^2+a^2}{\Xi}\,d\phi\right)^2, \label{kerrmetric}
  \end{split}\tag{B.2}
\end{align}
where $ \Delta_r = (r^2+a^2)(1+\frac{r^2}{l^2})-2mr$,
 $\Delta_\theta = 1-\frac{a^2}{l^2}\cos^2\theta$,
$\rho^2 =\ r^2+a^2\cos^2\theta$, and $\Xi = 1-\frac{a^2}{l^2}$.

For the model described by the Lagrangian \eqref{lagrangian23}, the black hole solution has been solved perturbatively around the AdS-Schwarzschild metric, up to $2$nd order in the couplings $\alpha_2$ and $\alpha_3$. The solution is written as
\begin{align}
ds^2=-f(r)dt^2+\frac{1}{g(r)}dr^2+r^2 d\Omega_2,\tag{B.3}
\end{align}
where
\begin{align}
& f(r) = 1-\frac{2 M}{r}-\frac{\Lambda r^2}{3} - \frac{64 \pi \alpha_3}{27 r^7}\left(-270 M^3  +108 \Lambda M^2 r^3\right. \notag \\
&\quad \left. +18 \Lambda^2 M r^6+\Lambda^3 r^9\right) + \frac{4096}{891 r^{13}} \left(-1058508 \pi^2 \alpha_2 \alpha_3 M^4 r^3 \right.\notag \\
&\quad \left.-304722 \pi^2 \alpha_2 \alpha_3 \Lambda M^3 r^6 +1010394 \pi^2 \alpha_2 \alpha_3 M^3 r^4\right. \notag \\
&\quad -21384 \pi^2 \alpha_2 \alpha_3 \Lambda^2 M^2 r^9 +144342 \pi^2 \alpha_2 \alpha_3 \Lambda M^2 r^7 \notag \\
&\quad -240570 \pi^2 \alpha_2 \alpha_3 M^2 r^5 +7107507 \pi^2 \alpha_3^2 M^5 \notag \\
&\quad \left.+57024 \pi^2 \alpha_3^2 \Lambda M^4 r^3 -6980904 \pi^2 \alpha_3^2 M^4 r\right. \notag \\
&\quad -462132 \pi^2 \alpha_3^2 \Lambda^2 M^3 r^6 +1011852 \pi^2 \alpha_3^2 \Lambda M^3 r^4 \notag \\
&\quad  +1679616 \pi^2 \alpha_3^2 M^3 r^2-46728 \pi^2 \alpha_3^2 \Lambda^3 M^2 r^9 \notag \\
&\quad \left.+288684 \pi^2 \alpha_3^2 \Lambda^2 M^2 r^7 -481140 \pi^2 \alpha_3^2 \Lambda M^2 r^5 \right. \notag \\
&\quad \left.-330 \pi^2 \alpha_3^2 \Lambda^4 M r^{12}-11 \pi^2 \alpha_3^2 \Lambda^5 r^{15} \right), \tag{B.4}
\end{align}
and
\begin{align}
 & g(r)= 1-\frac{2 M}{r}-\frac{\Lambda  r^2}{3}-\frac{64}{27 r^7}\left(2646 \pi  \alpha_3 M^3 \right. \notag \\
& \quad \left. +594 \pi  \alpha_3 \Lambda  M^2 r^3-1458 \pi  \alpha_3 M^2 r +18 \pi  \alpha_3 \Lambda ^2 M r^6  \right. \notag \\
&\quad \left.+\pi  \alpha_3 \Lambda ^3 r^9\right)-\frac{4096}{891 r^{13}} \left(-96228 \pi ^2 \alpha_2 \alpha_3 M^4 r^3 \right. \notag \\
& \quad \left.+16038 \pi ^2 \alpha_2 \alpha_3 \Lambda  M^3 r^6  +48114 \pi ^2 \alpha_2 \alpha_3 M^3 r^4\right. \notag \\
&\quad+5346 \pi ^2 \alpha_2 \alpha_3 \Lambda ^2 M^2 r^9-16038 \pi ^2 \alpha_2 \alpha_3 \Lambda  M^2 r^7 \notag \\
&\quad +1649241 \pi ^2 \alpha_3^2 M^5 +648648 \pi ^2 \alpha_3^2 \Lambda  M^4 r^3\notag \\
&\quad -997272 \pi ^2 \alpha_3^2 M^4 r +226908 \pi ^2 \alpha_3^2 \Lambda ^2 M^3 r^6 \notag \\
&\quad \left. -96228 \pi ^2 \alpha_3^2 \Lambda  M^3 r^4+32472 \pi ^2 \alpha_3^2 \Lambda ^3 M^2 r^9\right.\notag \\
&\quad \left.-80190 \pi ^2 \alpha_3^2 \Lambda ^2 M^2 r^7+330 \pi ^2 \alpha_3^2 \Lambda ^4 M r^{12}\right. \notag \\
&\quad \left.+11 \pi ^2 \alpha_3^2 \Lambda ^5 r^{15}\right).\tag{B.5}
\end{align}
The integration constants in solving the equation of motion have been chosen to fix $M$ as the canonical energy of the system. This choice simplifies the subsequent analysis of extended black hole thermodynamics. Otherwise, $M$ should be regarded as an ordinary mass parameter; it implies that the physical energy associated with the thermodynamics becomes a function of the mass parameter, $\Lambda$ and $\alpha_m$'s, requiring special attention to be paid to the integrability condition.

\section*{Appendix C}

In this Appendix, we study the extended black hole thermodynamics for the Lovelock gravity in $D=5$ dimensions, where only Gauss-Bonnet terms are involved. The relevant higher curvature terms are given by
\begin{align}
   \alpha\, F_{GB} =\frac{\alpha}{16 \pi}\left( R_{\mu\nu\rho\sigma} R^{\mu\nu\rho \sigma }-4 R_{\mu\nu} R^{\mu\nu}+R^2 \right). \tag{C.1}
\end{align}
The theory has an exact spherically  symmetric black hole solution with the metric \cite{Cai:2001dz}
\begin{align}
ds^2=-f(r)dt^2+\frac{1}{f(r)}dr^2+r^2 d\Omega_3,\tag{C.2}
\end{align}
where $f(r)=1+\frac{r^2} {4 \alpha }\left(1-\sqrt{1+\frac{64 \alpha  M}{3 \pi  r^4}-\frac{8 \alpha }{l^2}}\right)$.
According to our extended formalism, the extended first law of thermodynamics can be obtained from the formula
\begin{align}
    \tiny
   \int_{S_\infty}^{(reg.)}\!\left(\tilde{\delta} \mathbf{Q}_{\xi_H} \! -\!  \xi_H\! \cdot\! \boldsymbol{\Theta}[\tilde{\delta} \phi] \right)\! -\! T\tilde\delta S \!=\! V \tilde{\delta}P\!+V_{GB} \tilde{\delta} \alpha, \tag{C.3}
\end{align}
where $V_{GB}$ is  defined as a geometric quantity $V_{GB} \equiv  -\int_\mathcal{V_\infty}^{(reg.)} F_{GB} \, \xi_H \cdot \boldsymbol{\epsilon}$. The first term on the left hand side can be calculated explicitly by substituting the metric into it, which reduces to
\begin{align}
    \tilde\delta M - \frac{M}{\alpha}\left(1-\frac{1}{\sqrt{1-\frac{8 \alpha }{l^2}}}\right) \tilde\delta\alpha. \tag{C.4}
\end{align}
Putting it back into eq.(C.3),  we get the extended first law as
\begin{align}
\tilde\delta M=   T\tilde\delta S \!+\! V \tilde{\delta}P\!+(V_{GB}+\Delta V_{GB}) \tilde{\delta} \alpha, \tag{C.5}
\end{align}
where we have identified $\Delta V_{GB}\equiv\frac{M}{\alpha}\left(1-\frac{1}{\sqrt{1-\frac{8 \alpha }{l^2}}}\right)$ and $\Delta V=0$. Note that incorporating the Casimir energy into $M$ will slightly modify the concrete expression of the extended first law, which doesn't occur in the conventional thermodynamics.

Next we examine the correlation between $\Delta M\equiv M-M_K$ and $\Delta V_{GB}$, as proposed in eq.(30). In $D$ dimensions, the Komar mass formula is $M_K\equiv \frac{D-2}{D-3}  \int_{S_{\infty}}^{(reg.)} \mathbf{Q}_{\xi_t}$, which leads to $M_K=\frac{M}{\sqrt{1-\frac{8 \alpha }{l^2}}}$. Surely we find the relation $\Delta M=\alpha \Delta V_{GB}$. Notice eq.(30) should be generalized to $\Delta M =\sum_n\frac{2(n-1)}{D-3}\alpha_n \Delta V_n$ in $D$ dimensions.

The derived first law (C.5) can be directly verified by inserting the expressions of the thermodynamic quantities, or comparing with the existing literature \cite{Frassino:2014pha,Dolan:2014vba}.

\vfill
\end{document}